\documentclass[twocolumn,article,amsmath,amssymb,aps]{revtex4}
\usepackage{graphicx}
\usepackage{epsfig}
\usepackage{multirow}

\begin{document}
\title{Interplay of classical and quantal features within the coherent state model}
\author{A. A. Raduta$^{1,2}$, C. M. Raduta $^{1}$}
\affiliation{$^{1}$Institute of Physics and Nuclear Engineering$\mathrm{,}$ P.O. Box MG06, Bucharest 077125, Romania\\
$^{2}$Academy of Romanian Scientists, 54 Splaiul Independentei, Bucharest 050094, Romania}

\begin{abstract}
The classical and quantal features of a quadrupole coherent state and its projections
over angular momentum and boson number are quantitatively analyzed in terms of the departure of the Heisenberg uncertainty relations from the classical limit. This study is performed alternatively for two choices of the pairs of  conjugate coordinates. The role of deformation as mediator of classical and quantal behaviors is also commented. Although restoring the rotational and gauge symmetries makes the quantal features manifest dominantly for small deformation, these are blurred by increasing the deformation which pushes the system toward a classical limit.
\end{abstract}
\maketitle 

\renewcommand{\theequation}{1.\arabic{equation}}
\setcounter{equation}{0}
\label{sec:level1}
\section{Introduction}
Within a many body theory, defining the collective degrees of freedom is of paramount importance.
Moreover, the treatment of their motion and the coupling with non-collective coordinates is not an easy task. Some formalisms use a variational time dependent procedure which actually provides a semiclassical picture where the work is hopefully simplified \cite{BarVen,Vill}. The dequantization method is most reliable when the variational state is of a coherent type. Indeed, in that case quantizing the classical trajectories, the resulting spectrum might be close to that associated to the initial many body Hamiltonian. Such a treatment can be applied also to quadrupole boson Hamiltonians \cite{Rad95,Rad2010}.

The overcomplete property of a coherent state allows for accounting of the dynamics causing the collective motion. Indeed, expanding the coherent state in a Hilbert space basis, no expansion coefficient is vanishing. Due to this property, for a quadrupole boson Hamiltonian  contributions
in the whole boson space to its eigenvalues are included, which is not the case when a diagonalization procedure is adopted.
As a matter of fact this property was exploited when an axially deformed state of quadrupole type was used to generate through angular momentum projection a set of states which describe reasonable well the main properties of the rotational states from the ground band \cite{Rad76}. This idea was extended by considering two orthogonal polynomial excitations of the coherent state, each of them being orthogonal on the ground. By construction the orthogonality restriction of the three deformed states is preserved after the projection is performed. The additional two sets of projected states
have  properties  which are specific for the member states of the $\beta$ and $\gamma$ bands, respectively.
In the restricted boson state generated by the three sets of angular momentum projected states, an effective boson Hamiltonian is treated. The formalism, called the Coherent State Model (CSM), was successfully applied to describe the excitation energies and the $E2$ decay properties of a large number of nuclei \cite{Rad82}.

Here we try to answer a few questions related with the classical behavior of a boson system described by a state projected out from a coherent state. The coherent state has the property that minimizes the uncertainty relations, which in fact defines the classical character of the state.
The coherent state breaks several symmetries which can be restored by projection procedures.
We raise the question whether by restoring a symmetry the classical behavior is affected  or not.
The answer to this question is formulated in terms of the departure of the uncertainty relations from the classical limit.

The Heisenberg uncertainty relations are associated to a given pair of conjugate coordinates.
The second issue addressed in this work is whether the classical or non-classical features prevail depending on the chosen conjugate coordinates. 
Concluding, we aim at studying the effect of symmetry restoration on the Heisenberg uncertainty relations associated to two distinct pairs of conjugate coordinates.

The objective of this work will be described according to the following plan.
In Section II the basic properties of an axially deformed coherent state are presented. The uncertainty relations for the conjugate coordinates ($\hat{\alpha},\hat{\pi}$) are written for the unprojected, the angular momentum projected and the angular momentum plus the gauge symmetry projected states, in Section III. 
The same study but for the boson number operator and its conjugate phase, is achieved in Section IV. The numerical analysis presented in Section V, concerns the quantitative consideration of the equations derived in the previous sections. The final conclusion are drawn in Section VI.

\renewcommand{\theequation}{2.\arabic{equation}}
\setcounter{equation}{0}
\label{sec:level2}
\section{Axially deformed coherent state}

Let us consider the coherent state defined with the z-component of the quadrupole boson operators
$b^{\dagger}_{2\mu},b_{2\mu}$ with $-2\le\mu\le2$:
\begin{equation}
|\Psi\rangle =e^{(db^{\dagger}_{20}-d^*b_{20})}|0\rangle ,
\end{equation}
where $|0\rangle$ stands for the boson vacuum state while $d$ is a complex number. 
The coherent nature of this function is determined by:
\begin{equation}
b_{20}|\Psi\rangle =d|\Psi\rangle,\;\;\langle\Psi|b^{\dagger}_{20}=d^*.
\label{coherst}
\end{equation}
Using the Baker-Campbell-Hausdorff factorization:
\begin{equation}
e^{A+B}=e^Ae^Be^{-\frac{1}{2}[A,B]},
\end{equation}
the coherent state is written in the form:
\begin{equation}
|\Psi\rangle =e^{-\frac{|d|^2}{2}}e^{db^{\dagger}}|0\rangle.
\end{equation}
The average of the quadrupole operator:
\begin{equation}
Q_{20}=q_0\left(b^{\dagger}_{20}+b_{20}\right),
\end{equation}
with the coherent state has the expression:
\begin{equation}
\langle\Psi|Q_{20}|\Psi\rangle = 2q_0 Re\; d.
\label{quadrupole}
\end{equation}
The function $|\Psi\rangle$ is a vacuum state for the shifted quadrupole boson operator:
\begin{equation}
\left(b_{20}-d\right)|\Psi\rangle=0.
\end{equation}
Obviously the function $|\Psi\rangle$ has not a definite angular momentum, i.e. it is not eigenfunction of the angular momentum operator squared, $\hat{J}^2$. However it is eigenstate of $\hat{J}_z$. Due to this feature we say that $|\Psi\rangle$ is an axially deformed function.
Notice that Eq.(\ref{quadrupole}) asserts that $Re\; d$ plays the role of a deformation parameter.

For real $d$ the wave function $|\Psi\rangle$ has been used as generating function for the members of the ground state rotational bands \cite{Rad82,Haap,Rad76}. Moreover, constructing two mutually orthogonal states (as low order polynomial excitations of $|\Psi\rangle$), each of them being orthogonal on $|\Psi\rangle$, and requiring that the three intrinsic states satisfy a number of criteria suggested by the experimental data one generates, by angular momentum  projection, the member states of 
beta and gamma bands respectively. In the restricted boson space an effective boson Hamiltonian was defined. Actually these are the main ingredients of the Coherent State Model \cite{Rad82} which has been successfully used for describing in a realistic manner both the excitation energies and the $E2$ transition features.

Since we want to discuss the classical properties which might be described with the CSM we restrict our considerations to the case of real $d$.
\renewcommand{\theequation}{3.\arabic{equation}}
\setcounter{equation}{0}
\label{sec:level3}
\section{The quadrupole coordinate and its conjugate momentum}

\subsection{Unprojected state}

The conjugate coordinates:
\begin{eqnarray}
\hat{\alpha}_{20} &=&\frac{1}{\sqrt{2}}\left(b^{\dagger}_{20}+b_{20}\right),\nonumber\\
\hat{\pi}_{20} &=&\frac{i}{\sqrt{2}}\left(b^{\dagger}_{20}-b_{20}\right),
\end{eqnarray}
satisfy the equation:
\begin{equation}
[\hat{\alpha}_{20},\hat{\pi}_{20}]=i,
\end{equation}
where "$i$" denotes the imaginary unit.
Conventionally, we use the units system where $\hbar=1$.

The averages of $\hat{\alpha}$ and $\hat{\alpha}^2$ on $|\Psi\rangle$ are:
\begin{equation}
\langle \Psi|\hat{\alpha}_{20}|\Psi\rangle =\sqrt{2} d,\;\;
\langle \Psi|\hat{\alpha}^2_{20}|\Psi\rangle =2d^2+\frac{1}{2}.
\end{equation}
The conjugate momentum and its square have the averages:
\begin{equation}
\langle \Psi|\hat{\pi}_{20}|\Psi\rangle =0,\;\;
\langle \Psi|\hat{\pi}^2_{20}|\Psi\rangle =\frac{1}{2}.
\end{equation}
Using these results, the uncertainty relation associated to the conjugate coordinates $\alpha$ and $\pi$ has the form:
\begin{equation}
\Delta\hat{\alpha}_{20}\Delta\hat{\pi}_{20}=\frac{1}{2},
\end{equation}
where by $\Delta x$ one denotes the dispersion of the coordinate $x$. Notice that the dispersion product reaches the minimum value of the set allowed by the Heisenberg uncertainty principle.
Due to this feature one asserts that the coherent state $|\Psi\rangle$ is an optimal state to describe the properties which define the border of quantum and classical behavior.

As already mentioned, the coherent  state has not a definite angular momentum. The question is whether the classical features revealed by $|\Psi\rangle $ are preserved when the rotational symmetry is restored, i.e. from the deformed state one projects out the components of a definite angular momentum. A measure of the deviation from the classical behavior is again the departure of the dispersion product from the classical value. This will be analytically expressed in the next Section.

\subsection{Projected spherical states}
Through angular momentum projection one generates a set of orthogonal states:
\begin{equation}
\phi^{(g)}_{JM}=N^{(g)}_JP^J_{M0}|\Psi\rangle ,
\end{equation}
where $P^J_{MK}$ denotes the angular momentum projection operator
\begin{equation}
P^{J}_{MK}=\frac{2J+1}{8\pi^2}\int D^{J*}_{MK}(\Omega)\hat{R}(\Omega)d\Omega,
\end{equation} 
with $D^{J*}_{MK}$ standing for the Wigner function, or rotation matrix, $\hat{R}(\Omega)$
is the rotation defined by the Euler angles $\Omega$, while $N^{(g)}_J$ is the normalization factor.
The projected functions account for the main features of the rotational ground band \cite{Rad82}. For this reason the function is accompanied by the upper index $^{(g)}$.
The norms have been  analytically studied for any deformation and moreover very simple formulas for near vibrational and well deformed regimes have been obtained \cite{Rad012}. For the sake of completeness we give the necessary expressions:
\begin{equation}
\left(N^{(g)}_J\right)^{-2}=(2J+1)I^{(0)}_Je^{-d^2},
\end{equation}
with 
\begin{equation}
I^{(k)}_J(x)=\int^{1}_{0}P_J(y)\left(P_2(y)\right)^ke^{xP_2(y)}dy,\;\;x=d^2,
\end{equation}
where $P_k(x)$ are  the Legendre polynomial of rank $k$.
Expectation values of the conjugate coordinates and their squares have the expressions:
\begin{eqnarray}
&&\langle \phi^{(g)}_{JM}|\hat{\alpha}|\phi^{(g)}_{JM}\rangle =\sqrt{2}dC^{J\;2\;J}_{M\;0\;M}C^{J\;2\;J}_{0\;0\;0},
\nonumber\\
&&\langle \phi^{(g)}_{JM}|\hat{\alpha}^2|\phi^{(g)}_{JM}\rangle =\frac{1}{2}\nonumber\\
&+&d^2\left[\sum_{J'=0,2,4}C^{J\;J'\;J}_{M\;0\;M}C^{J\;J'\;J}_{0\;0\;\;0}\left(C^{2\;2\;J
'}_{0\;0\;0}\right)^2\right. \nonumber\\
&+&\left.\sum_{J'=0,2,4}\left(C^{J'\;2\;J}_{0\;0\;\;0}\right)^2\left(C^{J'\;2\;J}_{M\;0\;M}\right)^2
\left(\frac{N^{(g)}_{J}}{N^{(g)}_{J'}}\right)^2\right],\nonumber\\
&&\langle \phi^{(g)}_{JM}|\hat{\pi}|\phi^{(g)}_{JM}\rangle=0,\nonumber\\
&&\langle \phi^{(g)}_{JM}|\hat{\pi}^2|\phi^{(g)}_{JM}\rangle=\frac{1}{2}\nonumber\\
&+&
d^2\left[-\sum_{J'=0,2,4}C^{J\;J'\;J}_{M\;0\;M}C^{J\;J'\;J}_{0\;0\;0}\left(C^{2\;2\;J'}_{0\;0\;0}\right)^2\right. \nonumber\\
&+&\left.\sum_{J'=0,2,4}\left(C^{J'\;2\;J}_{0\;0\;0}\right)^2\left(C^{J'\;2\;J}_{M\;0\;M}\right)^2
\left(\frac{N^{(g)}_{J}}{N^{(g)}_{J'}}\right)^2\right].
\end{eqnarray}
Standard notation, $C^{j_1\;j_2\;j}_{m_1\;m_2\;m}$, for Clebsch-Gordan coefficients is used.
From here the dispersions of $\hat{\alpha}$ and $\hat{\pi}$ are readily obtained and then the dispersion product is analytically expressed.

\subsection{Simultaneous restoration of the rotation and gauge symmetries}
The boson number projection operator is:
\begin{equation}
\hat{P}_N=\frac{1}{2\pi}\int_{0}^{2\pi}e^{i\phi(\hat{N}-N)}d\phi.
\end{equation} 
Applying successively the projection operators $P^J_{MK}$ and $\hat{P}_N$ on the coherent state $\Psi$, one obtains a state of good angular momentum and boson number:
\begin{equation}
|NJM\rangle ={\cal N}_{NJ}\hat{P}_{N}P^{J}_{MK}|\Psi\rangle.
\end{equation}
Here ${\cal N}_{JN}$ denotes the normalization factor and has the expression:
\begin{equation}
\left({\cal N}_{NJ}\right)^{-2}=e^{-d^2}\frac{d^{2N}}{N!}(2J+1){\cal S}_{NJ},
\end{equation}
where the matrix ${\cal S}_{mJ}$ is defined by:
\begin{equation}
{\cal S}_{mJ}=\int_{0}^{1}\left(P_2(x)\right)^mP_{J}(x)dx.
\end{equation}
Following the path described in Ref. \cite{Rad76} one obtains:
\begin{equation}
{\cal S}_{lJ}(d)=\sum_{m=0}^{l}\frac{(-)^{l-m}3^m(l)!(2m)!(m+\frac{1}{2}J)!}{2^{l-J}m!(l-m)!(m-\frac{1}{2}J)!(2m+J+1)!}.
\end{equation}
The overlap matrix elements given above satisfy the restriction: they are nonvanishing only if $l\le J/2$.

The explicit expression of the projected state is:
\begin{equation}
|NJM\rangle = {\cal N}_{NJ}e^{-d^2/2}\frac{d^N}{N!}\frac{2J+1}{8\pi^2}\int D^{J*}_{M0}(\Omega)\hat{R}(\Omega)\left(b^{\dagger}_{20}\right)^N d\Omega|0\rangle,
\end{equation}
where $\hat{R}(\Omega)$ is the rotation defined by the set of Euler angles $\Omega$. 

The expectation values of the conjugate variables $\hat{\alpha}_{20}$ and $\hat{\pi}_{20}$ are equal to zero since each of the composing terms changes the boson number by one unit. Therefore, the corresponding dispersions squared are just the average values of their squares. By direct calculations one finds:
\begin{eqnarray}
&&\Delta\hat{\alpha}_{20}\Delta\hat{\pi}_{20}=\\
&&\frac{1}{2}+\sum_{J'=0,2,4}\left(C^{J'\;2\;J}_{M\;0\;M}\right)^2
\left(C^{J'\;2\;J}_{0\;0\;0}\right)^2d^2\left(\frac{{\cal N}_{NJ}}{{\cal N}_{(N-1)J'}}\right)^2.\nonumber
\end{eqnarray}

\renewcommand{\theequation}{4.\arabic{equation}}
\setcounter{equation}{0}
\label{sec:level4}
\section{Uncertainty relation for boson number and its conjugate phase}

\subsection{The case of intrinsic state}
In this section we consider again that $d$ is a complex number.
Let us denote by $\hat{N}_0$ the boson number operator:
\begin{equation}
\hat{N_0}=b^{\dagger}_{20}b_{20}.
\end{equation}
Writing the operator $\hat{N}^2_0$ in a normal order, the expectation values for $\hat{N}_0$ and 
$\hat{N}^2_0$ can be easily calculated:
\begin{eqnarray}
\langle\Psi|\hat{N}_0|\Psi\rangle&=&|d|^2\equiv N_0,\nonumber\\
\langle\Psi|\hat{N}^2_0|\Psi\rangle&=&|d|^2+|d|^4.
\end{eqnarray}
Thus the dispersion of the boson number operator can be easily calculated:
\begin{equation}
(\Delta \hat{N})^2=|d|^2\equiv N_0.
\label{DeltaN}
\end{equation}
Writing the complex number $d$ in the polar form 
\begin{equation}
d=|d|e^{i\varphi}
\end{equation}
and using Eq. (\ref{coherst}) 
one obtains:
\begin{eqnarray}
\langle \Psi|b_{20}|\Psi\rangle &=&|d|e^{i\varphi}=N_0^{1/2}e^{i\varphi},\nonumber\\
e^{i\varphi}&=& \langle \Psi|b_{20}|\Psi\rangle\left(\langle\Psi|\hat{N}_0|\Psi\rangle\right)^{-1/2}.
\end{eqnarray}
The question which arises is whether such a factorization holds also for operators whose averages are involved in the above equation.
Before dealing with the quantum mechanical problem of the boson number and its conjugate phase we would like to present first the classical counterpart of this long standing problem.

Let H be a Hamiltonian defined in terms of the quadrupole boson operators $b^{\dagger}_{2m},b_{2m}$ with $-2\le m\le 2$ and consider the Time Dependent Variational Principle ($TDVP$) equation:
\begin{equation}
\delta \int_{0}^{t} \langle \Psi|H-i\frac{\partial}{\partial t'}|\Psi\rangle dt'=0,
\label{varpr}
\end{equation}
where the variational state is the coherent state $|\Psi\rangle$ (\ref{coherst}), with $d$ a complex number depending on time.
The $TDVP$ leads to the Hamilton equations of motion for the classical coordinates $d$ and $d^*$.
\begin{eqnarray}
\frac{\partial {\cal H}}{\partial d}&=&-i\stackrel{\bullet}{d}^*,\nonumber\\
\frac{\partial {\cal H}}{\partial d^*}&=&i\stackrel{\bullet}{d},
\end{eqnarray}
where ${\cal H}$ denotes the average of $H$ with $|\Psi\rangle$, while $^{\bullet}$ indicates the  time  derivative. Changing the classical coordinates by the transformation 
\begin{equation}
(d,d^*)\rightarrow (r,\varphi),
\end{equation}
with $r=|d|^2$, the equations of motion become:
\begin{eqnarray}
\frac{\partial {\cal H}}{\partial r}&=&-\stackrel{\bullet}{\varphi},\nonumber\\
\frac{\partial {\cal H}}{\partial \varphi}&=&\stackrel{\bullet}{r}.
\end{eqnarray}
These equations suggest that the classical image (the average of $\hat{N}_0$ with $\Psi$) of the boson number operator and the phase $\varphi$ are indeed conjugate classical coordinates, namely $r$ is a classical coordinate and  $\varphi$ its conjugate momentum. One can check that their Poisson bracket is equal to unity. Certainly it would be desirable that a pair of Hermitian operators whose commutator is unity, exists such that their averages with $\Psi$ are just the canonical conjugate classical coordinates $r$ and $\varphi$. In what follows we devote some space to the issue just formulated. 
 
It is useful to introduce the off-diagonal operator
\begin{equation}
\hat{P}^{\dagger}_0=\hat{N}_0^{-1/2}b^{\dagger}_{20}
\end{equation}
which is the quantal counterpart of Eq. (4.5)
The operator $\hat{N}_0^{-1/2}$ is defined by the following equation:
\begin{equation}
\hat{N}_0^{-1/2}=\frac{1}{\sqrt{\pi}}
\int^{\infty}_{-\infty}
dx\exp\left(-x^2\hat{N}_0\right).
\label{nlamin1/2}
\end{equation}

The Hermitian conjugate operator $\hat{P}_0$, satisfies the commutation relation:
\begin{equation}
\left[\hat{P}_0,\hat{N}\right]=\hat{P}_0.
\label{comPN}
\end{equation}
The conjugate coordinate corresponding to the boson number operator is:
\begin{equation}
\hat{\Phi}_0=-i\ln\hat{P}_0.
\label{defPhi}
\end{equation}
Indeed, considering the power expansion of the $\ln$ function in terms of $(\hat{P}_0-1)$,  one checks that the operators $\hat{N}_0$ and $\hat{\Phi}_0$ satisfy
the commutation relation:
\begin{equation}
\left[\hat{N}_0,\hat{\Phi}_0\right]=i.
\label{nfieq1}
\end{equation}

For monopole bosons, the conjugate coordinates of boson number and phase were described in details
in Ref.\cite{Holevo}. By contradistinction to the monopole case here the rotation symmetry is broken. Indeed, while the boson number operator is a scalar, the phase operator $\hat{P}$ is a tensor of rank two and projection zero. It is an open question whether a construction of a scalar phase operator $\hat{P}$ is possible or not.

We note that our derivation of the phase operator is based on Eq.(\ref{comPN}). Actually this equation may be looked at as a defining equation for $\hat{P}_0$. Certainly the solution of this equation is not unique. For example, a possible solution is:
\begin{equation}
\hat{P}_0=b_{20}.
\end{equation}
In this case we have:
\begin{equation}
\hat{\Phi}_0=-i\ln b_{20}.
\end{equation}
One can check that the following equations for the expectation values hold:
\begin{equation}
\langle \Psi |\hat{\Phi}_0|\Psi\rangle =-i\ln d,\;\;\langle \Psi |\hat{\Phi}_0^2|\Psi \rangle =
-\left(\ln d\right)^2.
\end{equation}
Consequently, the corresponding dispersion is vanishing:
\begin{equation}
\Delta \hat{\Phi}_0=0,
\end{equation}
which reflects the fact that $\Psi$ is an eigenfunction of $\hat{\Phi}_0$:
\begin{equation}
\hat{\Phi}_0|\Psi\rangle =-i\left(\ln d\right)|\Psi\rangle .
\end{equation} 

A direct use of Eq.(\ref{defPhi}) to calculate the uncertainty relation for the boson number and phase, is quite a cumbersome task especially due to the logarithm function. However a considerable simplification is obtained by noticing that the deviation of $\hat{\Phi}_0$ from its expectation value can be expressed as:
\begin{equation}
\delta\hat{\Phi}_0=\frac{\delta\hat{P}_0}{\hat{P}_0}.
\end{equation}
We define a new dispersion of  the phase operator by:
\begin{equation}
D\hat{\Phi}_0=\frac{\Delta \hat{P}_0}{\langle \Psi|\hat{P}_0|\Psi\rangle}.
\end{equation}
We shall prove \cite{Holevo} that the newly defined quantity satisfy the Heisenberg uncertainty relation. Indeed, following the procedure of Ref.\cite{Carrut2} one successively obtains:
\begin{eqnarray}
\langle\Psi|\hat{P}_0|\Psi\rangle &=&de^{-|d|^2}\sum_{k=0}^{\infty}
\frac{|d|^{2k}}{k!\sqrt{k+1}},\nonumber\\
\langle\Psi|\hat{P}_0^2|\Psi\rangle &=&d^2e^{-|d|^2}\sum_{k=0}^{\infty}
\frac{|d|^{2k}}{k!\sqrt{(k+1)(k+2)}}.
\end{eqnarray}
 In the asymptotic region of $|d|$, compact forms  for the sums involved in the above equation were obtained in Ref. (\cite{Carrut}),  such that the final expressions for the considered expectation values are:
\begin{eqnarray}
\langle \Psi|\hat{P}_0|\Psi \rangle &=&\frac{d}{|d|}\left[1-\frac{1}{8|d|^2}+..\right],\nonumber\\
\langle \Psi|\hat{P}_0^2|\Psi\rangle &=&\frac{d^2}{|d|^2}\left[1-\frac{1}{2|d|^2}-\frac{3}{8|d|^4}+...\right].
\end{eqnarray}
With these results one finds that for large values of $|d|$, the following  uncertainty relation
holds:
\begin{equation}
\Delta\hat{N}_0D\hat{\Phi}_0=\frac{1}{2}.
\end{equation}
As we shall show in what follows, this equation is valid in the region of large number of bosons.
Indeed, in the region of large $|d|$ the $\Psi$ composing terms of maximal weights are those of large boson number.

The uncertainty relation of the boson number operator and its conjugate phase has been first studied for photons by Dirac 
\cite{Dirac27} and for oscillator by Susskind and Glover \cite{Suss}. The above equations have been obtained by representing the photon annihilation  operator as a product of a unitary operator, written as $U=e^{i\phi}$,  and a selfadjoint function of the boson number operator f(N). The solution is $f=\hat{N}^{1/2}$ and is based on the assumption that
$\phi$ is a selfadjoint operator. Later on it was proved that the conjugate phase variable is not well defined and therefore the corresponding uncertainty relation is doubtful. Indeed, one can check that the  operator U is not unitary and, consequently, $\phi$ is not a self adjoint operator and thereby cannot
be assigned to a physical observable \cite{Carrut}. The reason for non-unitary is the presence of a vanishing boson number in the spectrum of $\hat{N}$. Even if we exclude this value, which prevents $\hat{N}$ to be inversable, the phase is not well defined for small values of $N$ 
\cite{Louis}.
Indeed, denoting by $|n_0\rangle$ the eigenstates of $\hat{N}_0$, the matrix elements of Eq.
(\ref{nfieq1}) lead to:
\begin{equation}
\langle n_0|\hat{\Phi}_0|m_0\rangle=i\frac{\delta_{n_0,m_0}}{n_0-m_0},
\label{matrixfi}
\end{equation}
which doesn't make sense for small values of the boson number. However, for large values of the boson number this can be assimilated with a continuous variable and the ratio from the right hand side of Eq.(\ref{matrixfi}) is just the first derivative of the Dirac $\delta$-function
which is an well defined entity.

Positive attempts to define Hermitian operators depending on the phase, which together with the boson operator $\hat {N}_0$ satisfy the uncertainty relation, have been made by several authors
\cite{Carrut,Louis,Carrut2,Levine}. Thus, the operators:
\begin{eqnarray}
\hat{C}_0&=&\frac{1}{2}(\hat{P}_0+\hat{P}^{\dagger}_0),\nonumber\\
\hat{S}_0&=&\frac{1}{2i}(\hat{P}_0-\hat{P}^{\dagger}_0)
\end{eqnarray}
are Hermitian and satisfy the uncertainty relations \cite{Carrut2}:
\begin{eqnarray}
&&\Delta \hat{N}_0\Delta\hat{S}_0\ge \frac{1}{2}\langle \hat{C}_0\rangle,\nonumber\\
&&\Delta \hat{N}_0\Delta\hat{C}_0\ge \frac{1}{2}\langle \hat{S}_0\rangle ,
\end{eqnarray}
where $\langle ..\rangle$ denotes the average of the operator involved, with the coherent state.
The limitation on simultaneous measurement of observables $S_0$ and $C_0$ associated to the above mentioned Hermitian operators is expressed by the uncertainty product:
\begin{equation}
(\Delta \hat{S}_0)(\Delta \hat{C}_0)\ge\frac{1}{4}e^{-N_0},
\end{equation} 
with $N_0$ denoting the square of the dispersion $\Delta \hat{N}_0$.
A more symmetric uncertainty relation in the regime of large $|d|$ is obtained by combining the already obtained results:
\begin{equation}
(\Delta\hat{N}_0)^2\frac{(\Delta \hat{C}_0)^2+(\Delta \hat{S}_0)^2}{(\langle \hat{C}_0\rangle)^2+
(\langle \hat{S}_0\rangle)^2}\ge\frac{1}{4}.
\end{equation}
Although here we deal with quadrupole bosons the proof of the uncertainty relations mentioned above goes identically with those given in Refs.\cite{Carrut,Carrut2}.
\subsection{Dispersions of $\hat{N}$ and $\hat{P}$ on projected states}
 Note that while averaging the boson number operator
with the coherent state $|\Psi\rangle$ only the  component $b^{\dagger}_{20}b_{20}$ gives a non-vanishing contribution,  when the average is performed with the angular momentum projected state all the terms involved in the expression of the boson number operator, contribute. Therefore in this case the boson number $\hat{N}_0$ is to be replaced with the boson total number operator:
\begin{equation}
\hat{N}=\sum_{-2\le m\le 2}b^{\dagger}_{2m}b_{2m}.
\end{equation}
The phase operator $\hat{P}$ satisfying the commutation relation
\begin{equation}
\left[\hat{P},\hat{N}\right]=\hat{P},
\end{equation}
has the expression:
\begin{equation}
\hat{P}=\sum_{-2\le m\le 2}b_{2m} \hat{N}^{-1/2},
\end{equation}
where the reciprocal square root operator is defined as in Eq.\ref{nlamin1/2}.
Within the same spirit and with similar caution as before the conjugate phase operator is:
\begin{equation}
\hat{\Phi}=-i\ln \hat{P}.
\end{equation}

The expectation values of the boson number operator $\hat{N}$ and its square $\hat{N}^2$
have been analytically obtained in some  earlier publications of one of the authors (A.A.R.)\cite{Rad76}.
\begin{eqnarray}
\langle\phi^{(g)}_{JM}|\hat{N}|\phi^{(g)}_{JM}\rangle &=&|d|^2\frac{I^{(1)}_J}{I^{(0)}_J},\nonumber\\
\langle\phi^{(g)}_{JM}|\hat{N}^2|\phi^{(g)}_{JM}\rangle &=&|d|^2\frac{I^{(1)}_J}{I^{(0)}_J}+
|d|^4\frac{I^{(2)}_J}{I^{(0)}_J}.
\end{eqnarray}
One can check that the overlap integral ratios involved in the above equations are related by the following equation \cite{Rad82,Rad012}:
\begin{equation}
x^2\frac{I^{(2)}_J}{I^{(0)}_J}=\frac{1}{2}x(x-3)\frac{I^{(1)}_J}{I^{(0)}_J}+\frac{1}{4}\left(
2x^2+J(J+1)\right),\;\;x=|d|^2.
\end{equation}
From these equations one obtains the dispersion of $\hat{N}$:
\begin{eqnarray}
\left(\Delta \hat{N}\right)_{J}&=&-|d|^4\left(\frac{I^{(1)}_J}{I^{(0)}_J}\right)^2+\frac{1}{2}|d|^2(|d|^2-1)\frac{I^{(1)}_J}{I^{(0)}_J}\nonumber\\
&+&\frac{1}{4}\left(2|d|^4+J(J+1)\right).
\label{dn}
\end{eqnarray}

The uncertainty relations will be calculated by choosing as conjugate operator the phase operator 
$\hat{P}$ divided by its average value \cite{Holevo} and alternatively the Hermitian operators $\hat{C}$ and $\hat{S}$ defined as before \cite{Carrut2}:
\begin{eqnarray}
\hat{C}&=&\frac{1}{2}\left(\hat{P}+\hat{P}^{\dagger}\right),\nonumber\\
\hat{S}&=&\frac{1}{2i}\left(\hat{P}-\hat{P}^{\dagger}\right).
\end{eqnarray}

In what follows we shall describe a method of calculating the dispersion of the associated phase operator $\hat{P}$.
The average of $\hat{P}$ corresponding to the angular momentum projected state $|\Phi^{(g)}_{JM}\rangle$ is:
\begin{eqnarray}
&&\langle\phi^{(g)}_{JM}|\hat{P}|\phi^{(g)}_{JM}\rangle =
\left(N^{(g)}_{J}\right)^2
\langle\Psi|P^{J\dagger}_{M0}\sum_{\mu}b_{2\mu}P^J_{M0}\hat{N}^{-1/2}|\Psi\rangle\nonumber\\
&=&e^{-\frac{|d|^2}{2}}\left(N^{(g)}_{J}\right)^2\langle\Psi|P^{J\dagger}_{M0}\sum_{\mu}b_{2\mu}P^J_{M0}\nonumber\\
&\times&\frac{1}{\sqrt{\pi}}\int^{+\infty}_{-\infty}
\frac{(-)^mx^{2m}}{m!}\hat{N}^md^n\frac{b^{\dagger n}_{20}}{n!}dx|0\rangle\nonumber\\
&=&e^{-\frac{|d|^2}{2}}\left(N^{(g)}_{J}\right)^2\langle\Psi|P^{J\dagger}_{M0}\sum_{\mu}b_{2\mu}P^J_{M0}\frac{1}{\sqrt{n}}
d^n\frac{b^{\dagger n}_{20}}{n!}|0\rangle\nonumber\\
&=&e^{-\frac{d^2}{2}}C^{J\;2\;J}_{M\;0\;M}C^{J\;2\;J}_{0\;0\;0}\left(N^{(g)}_{J}\right)^2
\langle\Psi|P^J_{00}\nonumber\\
&\times&\sum_{n=1}^{\infty}\frac{d^n}{\sqrt{n}}\frac{\left(b^{\dagger}_{20}\right)^{n-1}}{(n-1)!}|0\rangle\nonumber\\
&=&de^{-|d|^2}C^{J\;2\;J}_{M\;0\;M}C^{J\;2\;J}_{0\;0\;0}\left(N^{(g)}_{J}\right)^2
\frac{2J+1}{2}\int_{-1}^{+1}dx\nonumber\\
&\times&\left[\sum_{m=0}^{\infty}\frac{1}{m!\sqrt{m+1}}\left(|d|^2P_2(x)\right)^mP_J(x)\right].
\end{eqnarray}
The final result is:
\begin{equation}
\langle \phi^{(g)}_{JM}|\hat{P}|\phi^{(g)}_{JM}\rangle =C^{J\;2\;J}_{M\;0\;M}C^{J\;2\;J}_{0\;0\;0}
d\frac{{\cal I}^{(0)}_{J}}{I^{(0)}_{J}},
\label{eqforP}
\end{equation}
where we denoted:
\begin{equation}
{\cal I}^{(0)}_{J}=\sum_{m=0}^{\infty}\frac{|d|^{2m}}{m!\sqrt{m+1}}{\cal S}_{mJ}.
\end{equation}
Applying a similar procedure as before but for $\hat{P}^2$, one obtains the final result:
\begin{equation}
\langle \phi^{(g)}_{JM}|\hat{P}^2|\phi^{(g)}_{JM}\rangle ={\cal C}^{J}_{M}d^2\frac{T^{(0)}_{J}}{I^{(0)}_{J}},
\label{eqforP2}
\end{equation}
with
\begin{eqnarray}
{\cal C}^{J}_{M}&=&\sum_{J'=0,2,4}C^{2\;2\;J'}_{0\;0\;0}C^{J\;J'\;J}_{0\;\;0\;\;\;0}C^{J\;J'\;J}_{M\;0\;M}\sum_{\mu}C^{2\; 2\; J'}_{\mu\;-\mu\;0},
\nonumber\\
T^{0}_J&=&\sum_{m=0}^{\infty}\frac{|d|^{2m}}{m!\sqrt{(m+1)(m+2)}}{\cal S}_{mJ}.
\end{eqnarray}

Having the expressions of the expectation values of $\hat{P}$ and $\hat{P}^2$, the dispersion of $P$ is readily obtained.
\begin{equation}
\left(\Delta\hat{P}\right)^2_{JM}=\langle \phi^{(g)}_{JM}|\hat{P}^2|\phi^{(g)}_{JM}\rangle
-\left(\langle \phi^{(g)}_{JM}|\hat{P}|\phi^{(g)}_{JM}\rangle\right)^2.
\end{equation}
Although calculating the average of $\hat{\Phi}$ is quite a cumbersome task, that is possible. However according to Ref.\cite{Holevo} the Heisenberg uncertainty inequality is satisfied by the 
dispersions of $\hat{N}$ and
\begin{equation}
(D\hat{P})_{JM}=\frac{(\Delta \hat{P})_{JM}}{|\langle  \phi^{(g)}_{JM}|\hat{P}|\phi^{(g)}_{JM}\rangle|}
\end{equation}
Therefore the departure from the classical limit is measured by 
$(\Delta{\hat{N}})_J(D\hat{P})_{JM}$.

\subsection{Uncertainty relations for $\hat{N}$ and $\hat{C}^2+\hat{S}^2$}
The expectation values of the Hermitian operators $\hat {C}$ and $\hat{S}$, defined by
\begin{eqnarray}
\hat{C}&=&\frac{1}{2}\left(\hat{P}+\hat{P}^{\dagger}\right),\nonumber\\
\hat{S}&=&\frac{1}{2i}\left(\hat{P}-\hat{P}^{\dagger}\right)
\end{eqnarray}
are easily obtained from Eqs.(\ref{eqforP}) and (\ref{eqforP2}).
\begin{eqnarray}
\langle \phi^{(g)}_{J0}|\hat{C}|\phi^{(g)}_{J0}\rangle&=&\left(C^{J\;2\;J}_{0\;0\;0}\right)^2
(Re\; d)\frac{{\cal I}^{(0)}_J}{J^{(0)}_J},\nonumber\\
\langle \phi^{(g)}_{J0}|\hat{S}|\phi^{(g)}_{J0}\rangle&=&\left(C^{J\;2\;J}_{0\;0\;0}\right)^2
(Im\; d)\frac{{\cal I}^{(0)}_J}{J^{(0)}_J}.
\end{eqnarray}
Following the procedure described in the previous subsection we obtain:
\begin{eqnarray}
&&\langle \phi^{(g)}_{J0}|\hat{C}^2+\hat{S}^2|\phi^{(g)}_{J0}\rangle =\nonumber\\
&&|d|^2\sum_{J'}\left(C^{J\;2\;J'}_{0\;0\;0}\right)^2\frac{{\cal U}^{(0)}_{J'}}{I^{(0)}_{J}}+
\frac{5}{2}\frac{{\cal U}^{(0)}_{J}}{I^{(0)}_{J}},
\end{eqnarray}
with
\begin{equation}
{\cal U}^{(0)}_{J}=\sum_{k=0}^{\infty}\frac{|d|^{2k}}{(k+1)!}S_{kJ}.
\end{equation}

The normalized sum of dispersions associated to the two observables $\hat{C}$ and $\hat{C}$ is:
\begin{eqnarray}
&&\frac{(\Delta\hat{C})_{J0}^2+(\Delta\hat{S})_{J0}^2}{\langle\hat{C}\rangle_{J0}^2+\langle\hat{S}\rangle_{J0}^2}=\frac{1}{|d|^2\left(C^{J\;2\;J}_{0\;0\;0}\right)^4}\nonumber\\
&\times&\left[|d|^2\sum_{J'}\left(C^{J\;2\;J'}_{0\;0\;0}\right)^2\frac{{\cal U}^{(0)}_{J'}I^{(0)}_J}
{({\cal I}^{(0)}_J)^2}+\frac{5}{2}\frac{{\cal U}^{(0)}_{J}I^{(0)}_J}
{({\cal I}^{(0)}_J)^2}\right]-1\nonumber\\
&&\equiv(\Delta R)_J^2,
\label{dc2+ds2}
\end{eqnarray} 
where the low index $J0$ suggests that the involved dispersions and average values correspond to the angular momentum projected state $\phi^{(g)}_{J0}$. Also the notation
$\langle \hat{O}\rangle_{J0}$ was used for the average value of $\hat{O}$ with the mentioned projected state.

The uncertainty relation associated to the two observables is obtained by equating
\begin{equation}
F_J=(\Delta\hat{N})_J\sqrt{\frac{(\Delta\hat{C})_{J0}^2+(\Delta\hat{S})_{J0}^2}{\langle\hat{C}\rangle_{J0}^2+\langle\hat{S}\rangle_{J0}^2}}
\label{Fdef}
\end{equation}
to the product of the right hand sides of equation (\ref{dn}) and $(\Delta R)_J$ given by
 \ref{dc2+ds2}.
The departure of $F$ from the value of $1/2$ constitutes a measure for the quantal nature of the system behavior.

\section{Numerical analysis}
\renewcommand{\theequation}{5.\arabic{equation}}
\setcounter{equation}{0}
\label{sec:level5}
We start by giving the expansion weights of $\Psi$ corresponding to various boson basis:
\begin{eqnarray}
|\Psi\rangle &=&\sum_n C_n |n\rangle,\nonumber\\
|\Psi\rangle &=&\sum_{J}C_{J0}|J0\rangle,\nonumber\\
|\Psi\rangle &=&\sum_{NJ}C_{NJ0}|NJ0\rangle.
\end{eqnarray}
were $|n\rangle$ are eigenstates of the boson number operator $b^{\dagger}_{20}b_{20}$, $|J0\rangle$ denotes the eigenstates of angular momentum square, $\hat{J}^2$, and its projection on z-axis, $J_z$. The third basis $\{|NJ0\rangle\}$ is determined by the quantum numbers: the  boson number $N$, the angular momentum $J$ and z-projection of the angular momentum, 0.

Using the results described before, one finds out that the expansion weights have the following analytical expressions:
\begin{eqnarray}
C_{n}&=&e^{-d^2/2}\frac{d^n}{\sqrt{n!}},\nonumber\\
C_{J0}&=&\left(N^{(g)}_{J}\right)^{-1},\nonumber\\
C_{NJ}&=&\left({\cal N}_{NJ}\right)^{-1}.
\end{eqnarray}  
\begin{figure}
\begin{center}
\includegraphics[width=7cm]{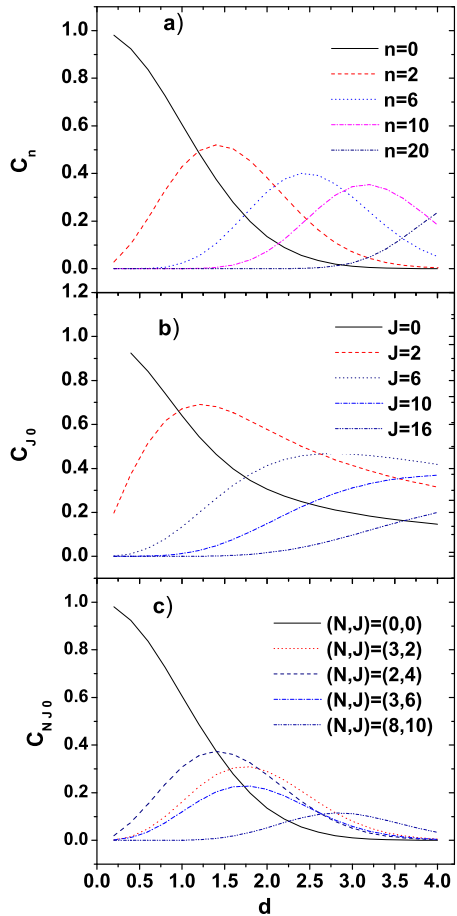}
\end{center}
\caption{(Color online) Expansion coefficients of the coherent state $|\Psi\rangle$ in three distinct basis,
$|n\rangle$ (panel a)), $|J0\rangle$ (panel b), and $|NJ0\rangle$ (panel c), are plotted as function of the deformation parameter $d$.}
\end{figure}
These weights have been plotted in Figs 1a)-1c), as function of the deformation parameter $d$.
From here we notice that for small $d$ some weights are vanishing which means that the corresponding states are missing in the expansion. The curves have maxima's for some deformations
which indicate that for such deformations the corresponding states, showing up in the expansion,
are the dominant components.
\begin{figure}
\begin{center}
\includegraphics[width=7cm]{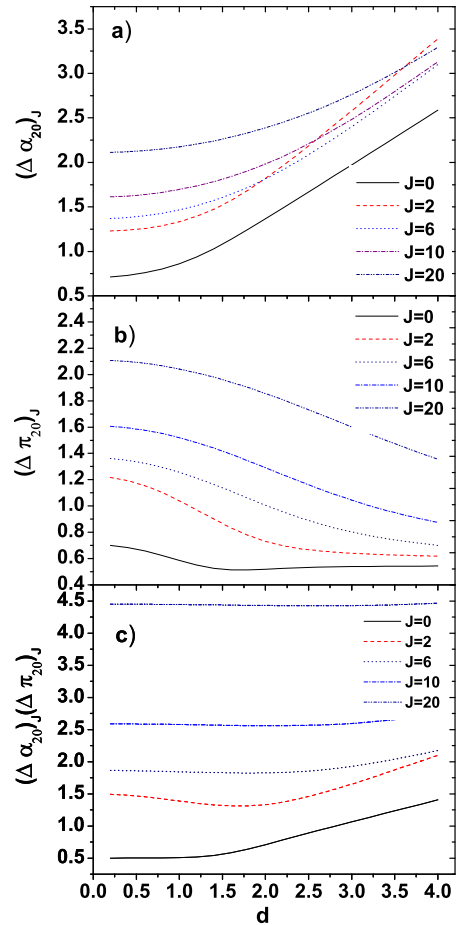}
\end{center}
\caption{(Color online) Dispersions of the conjugate coordinates $\alpha_{20}$ and $\pi_{20}$, are given in panels a) and b) respectively as functions of $d$. The product of the two dispersions is represented as function of $d$ in panel c). The three mentioned quantities correspond to the angular momentum projected states.}
\end{figure}
The dispersions product  of the conjugate coordinates $\hat{\alpha}_{20}$ and $\hat{\pi}_{20}$
calculated with the $J-$projected states is presented as function of $d$ in Fig. 2c). It is well known that the classical limit of this quantity is 1/2 (units of $\hbar$).
According to Fig. 2c) the $J=0$ projected state is the only projected state which behaves semiclassically in the region of small $d$ ($\le 1.5$). The remaining states lay apart from the classical limit. The larger the $J$, the larger the deviation from the classical limit. In the region of large $d (>3.)$ the deviation from the classical limit is an increasing function of $d$, irrespective
the values of $J$. This behavior trend is determined by the increasing nature of the dispersion of $\hat{\alpha}_{20}$. Thus for the both conjugate coordinates mentioned above, the nuclear deformation favor the quantal behavior of the system.

As we have already mentioned $\Psi$ breaks two symmetries, the gauge and rotational symmetries. So far we analyzed the wave function which restores the rotational symmetry. 
\begin{figure}[h!]
\begin{center}
\includegraphics[width=7cm]{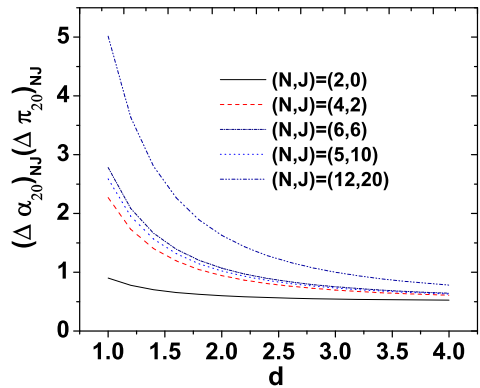}
\end{center}
\caption{(Color online) Dispersion product for the conjugate coordinates $\hat{\alpha}_{20}$ and $\hat{\pi}_{20}$,
corresponding to the $NJ$-projected states are, plotted as function of the deformation parameter $d$. }
\end{figure}
In Fig. 3 we represented the product of $\hat{\alpha}_{20}$ and $\hat{\pi}_{20}$ dispersions as function of $d$, using the states $|NJM\rangle$, which restore both symmetries mentioned above.
As seen in Fig. 3, the dispersion product has a strong $J$-dependence for small values of $d$
while for large valued of $d$, i.e. in the rotational limit, this tends to the classical limit.
In contrast to the case of the $J$-projected function here increasing $d$, favors the classical behavior. This is reflected in the decreasing $J$-dependence of the dispersion product as well as in approaching the classical value.

It is worth addressing the question whether the features mentioned above depend on the chosen pair  of conjugate coordinates. We attempt at giving not a general answer but analyze, for comparison, what happens when the pair of conjugate variables is
 ($\hat{N},\hat{P}$). The results are pictured in Figs. 4a)-4c). Dispersion of $N$ increases with $d$ and the split due to the $J$ dependence increases slowly from zero within a narrow interval.
Note that for $d$ going to zero the projected function $\phi^{(g)}_{JM}$ goes to the state
$|\frac{J}{2},\frac{J}{2},0,J,M\rangle$ \cite{Rad82} with the standard notation 
$|N,v,\lambda, J,M\rangle$: $N$ being the number of bosons, $v$ the seniority, $\lambda$ the missing quantum number, $J$ the angular momentum and $M$ the projection on the laboratory z-axis. Therefore, in the spherical limit $N$ becomes a good quantum number and the dispersion is vanishing.
By contrast, the normalized dispersion $D(P)$ has a large spread over $J$ for small values of $d$
but for large deformation the $J\ne 2$ dispersions attain a common value.
Note that apart from the quantitative aspects, the dispersion product preserves the look of $\Delta N$.
\begin{figure}[h!]
\begin{center}
\includegraphics[width=7cm]{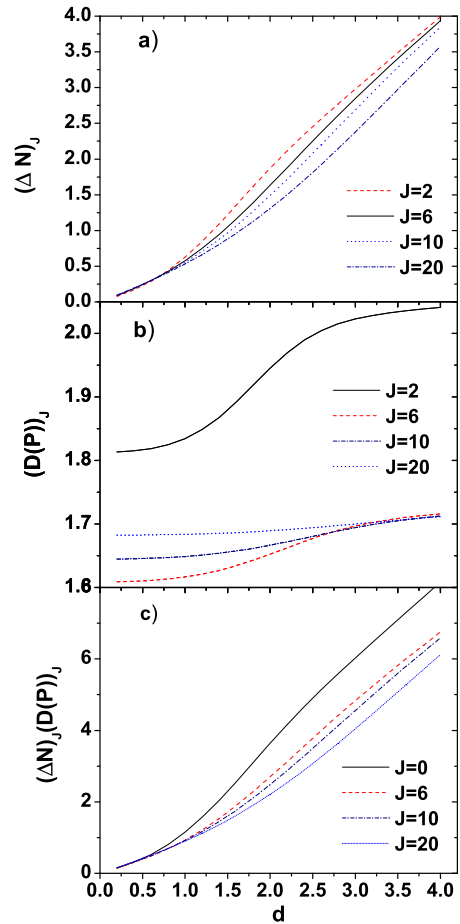}
\end{center}
\caption{(Color online) Dispersions for the boson number operator (panel a)) and the corresponding phase operator $P$ (panel b)) are plotted as function of $d$. Also, the dispersion product is given in panel c). Calculations are performed for angular momentum projected states.}
\end{figure}

\begin{figure}[h!]
\begin{center}
\includegraphics[width=7cm]{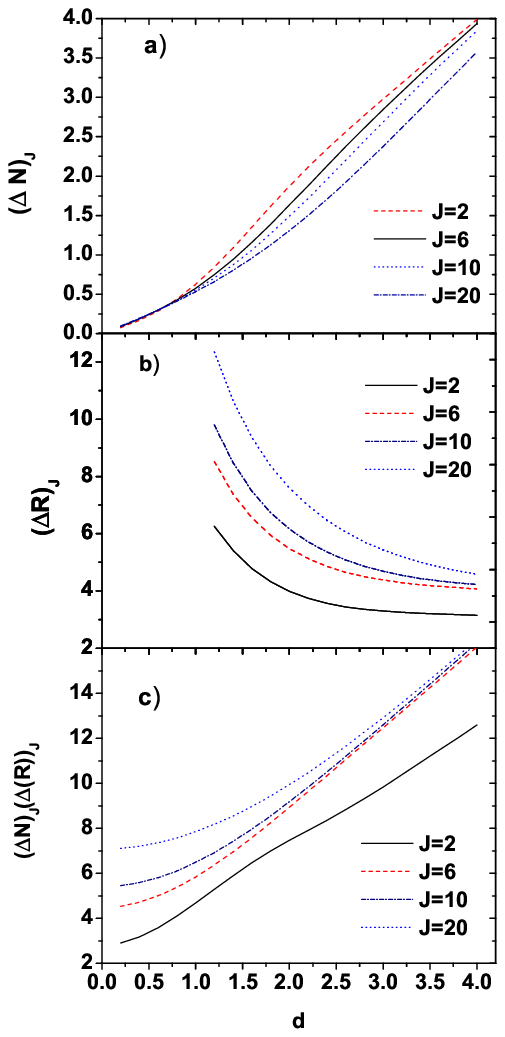}
\end{center}
\caption{(Color online) Dispersions for the boson number operator (panel a)) and for  the observable $\hat{R}$ defined by \ref{dc2+ds2} (panel b)) are plotted as function of $d$. The dispersions product is given in panel c). Calculations are performed for angular momentum projected states. The angular momenta are those specified in the figure legend.}
\end{figure}
The behavior of the pair of coordinates $(N,R)$ is visualized in Figs 5a)-5c). The split of R-dispersions due to their J-dependence is quite large for small deformation and is decreasing with $d$. The situation when dispersion of $R_J$ with $J\ne 2$ get a common value is reached for  $d$ larger than the maximum value shown in Fig. 4. The uncertainty relation for the pair $(N,R)$ is shown in the plot of $F_J$ (\ref{Fdef}) as a function of d. One notices that the departure from the classical limit is an increasing function of angular momentum. Also this is increasing with the nuclear deformation. It is worth noticing that for large deformation, the $J\ne 2$ values become indistinguishable.

\section{Conclusions}
\renewcommand{\theequation}{6.\arabic{equation}}
\setcounter{equation}{0}
\label{sec:level6}
Here we summarize the results obtained in the previous sections.
The expansion weights of the coherent states in three distinct bases exhibit a maximum when is represented as function of the deformation parameter $d$. The larger are the selected quantum numbers the larger is the deformation for which the weight is maximum.

In the ($\alpha,\pi $) representation only the $J=0$ projected state behaves classically and that happens for small values of $d$. In the region of  large $d$ the departure from the classical
picture is slightly increasing with the deformation.

The behavior of the ($\alpha,\pi$) pair of conjugate coordinates in a $NJ$-projected state is different from that described above for a $J$-projected state. Indeed, from Fig. 3 we notice that the quantal features prevail for small deformation, while for the rotational limit of large $d$
the associated Heisenberg relation approaches  the classical limit. Also, in this limit  the $J$
-dependence of the uncertainty relations is very weak.

From Figs. 2c) and 3 we conclude that in the $(\alpha, \pi)$ and J-projected states representation the system departs from the classical picture by increasing $d$ while for the $(N,J)$ projected states the larger the deformation the closer is the system to the classical behavior. In both cases the small deformation region is characterized by a quantal behavior reflected by the departure from the classical limits as well as by the split of the dispersion product due to the J-dependence.
Comparing the figures referring to the uncertainty relations for the $(\alpha, \pi)$ in the J projected and $NJ$  projected states respectively, one may conclude that the share of classical and quantal features depends on the symmetry of the wave function. In the specific situations the more symmetric is the system the closer is its behavior to the classical picture.

The delicate problem of boson number and the conjugate phase was treated by two alternative choices for the conjugate phase-like operator. In the first case the operator is ${\hat P}$ with a proper normalization. Although this is not a Hermitian operator the Heisenberg uncertainty relation holds for a large number of bosons. The dispersion product is quickly increasing with $d$ starting with values close to zero ($\approx 0.162$).The behavior for small deformation is justified by the fact that the $J$-projected state becomes eigenstate of $\hat{N}$. For larger $d$
the system behaves in a classical manner while for large deformation the quantal features prevail.
The split of the dispersion product due to its $J$-dependence is not significant for $J\ne 2$. One may say that although the departure from the classical limit of the dispersion product is large the classical feature reclaiming a weak $J$-dependence, still persists.
The dispersion product is increasing with $d$ and is also $J$ independent for large values of $J$. 

For the second alternative situation the phase like operators ${\hat C}$ and ${\hat S}$  were used to define, for the sake of having a symmetrical form, the dispersion of the observable $R$. The dispersion product denoted by $F_J$
is increasing with d. For small $d$ the split over J is large while for large deformation the values of $F_J$ for large $J$ are more or less the same.
Here, as well as in the case of $(\alpha,\pi)$ coordinates, the coordinate dispersion is increasing with $d$, while the conjugate momentum dispersion is decreasing when $d$ increases.

Comparing the results for $(\alpha,\pi)$ and $(N,P)$ /or $(N,R)$ coordinates we notice that the interplay of quantal and classical feature depends on the pair of conjugate coordinates under study.

Before the symmetries were restored the system behaves classically which is reflected by that the uncertainty relations achieve their minima, irrespective of the chosen pair of conjugate coordinate. Moreover, the expectation value for angular momentum square has a continue value 
\cite{Rad82}:
\begin{equation}
\langle \Psi|\hat{J}^2|\Psi\rangle = 6|d|^2.
\end{equation}
Symmetry projection leads to a J (or NJ-)-dependence for the uncertainty relations which is large for small deformation. Increasing $|d|$, the system tends to recover the classical behavior.

The main conclusions of our investigation are as follows:
\begin{itemize}
\item{In the intrinsic frame, where the unprojected states are used, the system behaves classically
irrespective of the chosen conjugate variables.}
\item{Also, angular momentum is a classical continuous variable as shown by Eq. (6.53).}
\item{In the projected basis $|NJM\rangle$ the dispersion product for $\alpha_{20}$
and $\pi_{20}$ reclaims a classical behavior for large deformation where the classical limit is approached for any value of J}
\item{Regarding the pair of coordinates $(N,R)$, as shown in Fig. 5, the system behaves classically only in high spin states, although the dispersion product is far from the classical value. However for large deformation and high spin the $J$ dependence vanishes.}
\item{For small deformation the quantal features prevail for both pairs of conjugate coordinates
considered.}
\item{Restoring the rotational and the gauge symmetries is a pure quantum mechanical operation which results in having a pronounced quantal behavior for small deformation.}
\item{Keeping in mind that the deformation itself is a classical variable, increasing its value
the quantal features are diminished in favor of the classical ones. In that respect one could say that the deformation plays the role of a mediator of classical and quantal behaviors.}
\item{The share of classical and quantal features depends on the symmetry of the wave function. In the specific situations considered here, the more symmetric is the system the closer is its behavior to the classical picture, in a large deformation regime.}
\item{Finally we may say that the present  work completes our previous studies upon the projected states of the coherent state $|\Psi\rangle$.}
\end{itemize}

{\bf Acknowledgment.} This work was supported by the Romanian Ministry for Education Research Youth and Sport through the CNCSIS project ID-2/5.10.2011. Also, A.A.R wants to thank 
Prof. Dr. M. Krivoruchenko for providing some relevant references concerning the number and phase uncertainty relation as well as for making interesting remarks about the preliminary version of the present paper.


\begin{references}
\bibitem{BarVen} M. Baranger and M. Veneroni, Ann. Phys. (NY) {\bf 114}, 123 (1978).
\bibitem{Vill} F. Villars, Nucl. Phys. {\bf A285}, 269 (1977).
\bibitem{Rad95} A. A. Raduta, V. Baran and D. S. Delion, Nucl. Phys. {\bf A588}, 431 (1995).
\bibitem{Rad2010} A. A. Raduta, R. Budaca and Amand Faessler, Jour. Phys. G: Nucl. Part. Phys.
{\bf 38}, 055102 (2011).
\bibitem{Rad76}A. A. Raduta and R. M. Dreizler, Nucl. Phys. {\bf A258}, 109 (1976).
\bibitem{Rad82} A. A. Raduta, V. Ceausescu, A. Gheorghe and R. M. Dreizler, Nucl. Phys. 
{\bf A381}, 253 (1982).
\bibitem{Haap} P. Haapakoski, T. Honkaranta and P. O. Lipas, Phys. Lett. {\bf 31 B}, 493 (1970).
\bibitem{Holevo}A.S. Holevo, {\it Probabilistic and Statistical Aspects of Quantum Theory,} 2nd edition, Edizioni della Normale, Pisa, 2011, ISBN: 978-88-7642-375-8 Nauka, Moscow, 1980, Russian translation, pp. 204-211.
\bibitem{Carrut2}P. Carruthers and M. M. Nieto,  Phys. Rev. Lett. {\bf 14}, 387 (1965).
\bibitem{Carrut}P. Carruthers and Michael Martin Nieto, Rev. Mod. Phys. {\bf 40}, 411 (1968).
\bibitem{Dirac27} P. A. M. Dirac, Proc. Roy. Soc. (London) {\bf A114}, 243 (1927).
\bibitem{Suss} L. Susskind and J. Glogow, Physics {\bf 1}, 49 (1964).
\bibitem{Louis} W. H. Louisell, Phys. Lett. {\bf 7}, 60, (1963).
\bibitem{Levine} R. D. Levine, The Journal of Chemical Physics, {\bf 44}, 3597 (1965).
\bibitem{Rad012} A. A. Raduta, R. Budaca and Amand Faessler, Ann. Phys.[NY] {\bf 327}, 671 (2012).
\end{references}
\end{document}